\documentclass[aps,floatfix,superscriptaddress,showpacs,showkeys]{revtex4}
\usepackage{amssymb,amsmath,amsfonts,latexsym,graphicx,epsfig}
\usepackage{color} 
\usepackage{titlesec}
\titleformat{\section}{\large\bfseries}{\thesection}{1em}{}

\newcommand{\bea}{\begin{eqnarray}}
\newcommand{\ena}{\end{eqnarray}}
\newcommand{\nn}{\nonumber\\}
\newcommand{\be}{\begin{equation}}
\newcommand{\en}{\end{equation}}

\newcommand{\la}{\langle}
\newcommand{\ra}{\rangle}

\begin{document}

\hfill DSF-2014-2 (Napoli),  MITP/14-080 (Mainz) 

\title{Heavy-to-light semileptonic decays of 
$\Lambda_b$ and $\Lambda_c$ baryons \\
in the covariant confined quark model} 

\author{Thomas Gutsche}
\affiliation{
Institut f\"ur Theoretische Physik, Universit\"at T\"ubingen,\\
Kepler Center for Astro and Particle Physics,\\
Auf der Morgenstelle 14, D-72076, T\"ubingen, Germany}
\author{Mikhail A. Ivanov} 
\affiliation{Bogoliubov Laboratory of Theoretical Physics, \\
Joint Institute for Nuclear Research, 141980 Dubna, Russia} 
\author{J\"{u}rgen G. K\"{o}rner} 
\affiliation{PRISMA Cluster of Excellence, Institut f\"{u}r Physik, 
Johannes Gutenberg-Universit\"{a}t, 
D-55099 Mainz, Germany} 
\author{Valery E. Lyubovitskij} 
\affiliation{
Institut f\"ur Theoretische Physik, Universit\"at T\"ubingen,\\
Kepler Center for Astro and Particle Physics,\\
Auf der Morgenstelle 14, D-72076, T\"ubingen, Germany}
\affiliation{ 
Department of Physics, Tomsk State University,  
634050 Tomsk, Russia} 
\affiliation{Mathematical Physics Department, 
Tomsk Polytechnic University,\\
Lenin Avenue 30, 634050 Tomsk, Russia} 
\author{Pietro Santorelli} 
\affiliation{
Dipartimento di Scienze Fisiche, Universit\`a di Napoli
Federico II, Complesso Universitario di Monte Sant' Angelo,
Via Cintia, Edificio 6, 80126 Napoli, Italy} 
\affiliation{
Istituto Nazionale di Fisica Nucleare, Sezione di Napoli, 
80126 Napoli, Italy}

\today

\begin{abstract}
We present a detailed analysis of the heavy-to-light semileptonic decays 
of the $\Lambda_b$ and $\Lambda_c$ baryons 
$\Lambda_b \to p \ell^- \bar\nu_\ell$ and $\Lambda_c \to n \ell^+ \nu_\ell$ 
in the covariant confined quark model. We calculate the invariant and  
helicity amplitudes of the two processes which are then used to analyze    
their angular decay distributions, their rates and asymmetry 
parameters. 
\end{abstract}


\maketitle

\section{Introduction}

Heavy-to-light semileptonic decays of heavy baryons are important 
physical processes for the determination of the Cabibbo-Kobayashi-Maskawa 
(CKM) matrix elements. In particular, a study of the exclusive decay 
$\Lambda_b \to p \mu^- \bar\nu_e$ at the Large Hadron Collider (LHC) 
affords the opportunity to determine the CKM matrix element $|V_{ub}|$. 
A discrepancy between the extractions of $|V_{ub}|$ from the exclusive and 
inclusive semileptonic $B$ meson decays at the $B$ factories~\cite{pdg12} 
is a long-standing puzzle in the heavy flavor sector of the Standard Model. 
Presently, the Particle Data Group~\cite{pdg12} reports the following 
averaged values for $|V_{ub}|$ 
\bea 
|V_{ub}|_{_{\rm excl.}} = \Big(4.41 \pm 0.15^{+0.15}_{-0.17}\Big) \times 
10^{-3}\,, \quad 
|V_{ub}|_{_{\rm incl.}} = (3.23 \pm 0.3) \times 10^{-3}\,.
\ena 
The exclusive result for $|V_{ub}|$ was extracted from using data  
from the Belle~\cite{Belle} and {\it BABAR}~\cite{BABAR} Collaborations  
for the semileptonic $\bar B \to \pi^+ \ell^- \bar\nu_\ell$ decay rate 
together with calculations for the $B \to \pi$ transition form factors 
in lattice QCD~\cite{Lattice_Vub}. Compared to the $B$ meson 
semileptonic decays the baryon transition $\Lambda_b \to p$ has 
an edge over the meson decay because the final state proton has a very 
distinct experimental
signature. It is therefore important to provide a thorough theoretical
decay analysis of the decay $\Lambda_b \to p \ell^- \bar\nu_\ell$ starting 
from a determination of the vector and axial form 
factors describing the current-induced $\Lambda_b \to p $ transition matrix
element. The calculation of the $\Lambda_b \to p$ form factors has 
been performed before using different versions of QCD sum 
rules~\cite{Huang:1998rq}-\cite{Khodjamirian:2011jp}, 
quark models~\cite{Datta:1995mv}-\cite{Wei:2009np} 
and lattice QCD~\cite{Detmold:2013nia}.  
In this paper we present calculations for the form factors characterizing  
the $\Lambda_b \to p \ell^- \bar\nu_\ell$ and 
$\Lambda_c \to n \ell^+ \nu_\ell$ transitions 
covariant confined quark model~\cite{Branz:2009cd}-\cite{Gutsche:2013oea}.

\section{
\boldmath{$\Lambda_b \to p \ell^- \bar\nu_\ell$} and 
\boldmath{$\Lambda_c \to n \ell^+ \nu_\ell$} matrix elements 
and observables} 

The effective Fermi Lagrangian for the semileptonic transitions 
$b\to u \ell^- \bar\nu_\ell$ and $c\to d \ell^+ \nu_\ell$ reads 
\be 
{\cal L}_{\rm eff} = \frac{G_F}{\sqrt{2}} \, 
\biggl[ 
V_{ub} \, (\bar u^a O_\mu b^a) \, (\bar\ell O^\mu \nu_\ell) 
\,+\, 
V_{cd} \, (\bar c^a O_\mu d^a) \, (\bar\ell O^\mu \nu_\ell) \biggr] 
\,+\, {\rm H.c.} 
\en
where $O_\mu = \gamma_\mu (1-\gamma_5)$ and  
$V_{qq'}$ are the Kobayashi-Maskawa matrix elements 
($|V_{ub}| =  0.00389$, $|V_{cd}| =  0.230$). 

This Lagrangian generates transitions on the quark level which in turn 
determine the heavy-to-light baryon transition matrix elements.  
The corresponding matrix elements of the exclusive transition
$\Lambda_b\to p \ell^- \bar\nu_\ell$ and 
$\Lambda_c\to n \ell^+ \nu_\ell$ are defined by 
\bea\label{eq:matr_LQN} 
M(\Lambda_b\to p  \ell^- \bar\nu_\ell) &=& 
\frac{G_F}{\sqrt{2}} \, V_{ub} \, 
\, \la p | \bar u O_\mu b | \Lambda_b \ra \, j^\mu_\ell \,,
\nonumber\\
M(\Lambda_c\to n \ell^+ \nu_\ell) &=&
\frac{G_F}{\sqrt{2}} \, V_{cd}^\ast \, 
\, \la n | \bar d O_\mu c | \Lambda_c \ra \, j^\mu_\ell \,,
\ena 
where $j^\mu_\ell$ is the leptonic current formed by the 
corresponding charged lepton and (anti) neutrino. 

The hadronic matrix elements 
$\la p | \bar u O_\mu b | \Lambda_{b} \ra$ and 
$\la n | \bar d O_\mu c | \Lambda_{c} \ra$ in~(\ref{eq:matr_LQN}) 
can be written in terms of six dimensionless, invariant form factors 
$f_i^{J}$ ($i=1, 2, 3$ and $J = V, A$), viz. 
\bea
\label{ffexpansion}
\la B_2\,|\,\bar s\, \gamma^\mu\, b\,| B_1 \ra &=&
\bar u_2(p_2)
\Big[ f^V_1(q^2) \gamma^\mu - f^V_2(q^2) i\sigma^{\mu q}/M_1
     + f^V_3(q^2) q^\mu/M_1 \Big] u_1(p_1)\,,
\nn
\la B_2\,|\,\bar s\, \gamma^\mu\gamma^5\, b\,| B_1 \ra &=&
\bar u_2(p_2)
\Big[ f^A_1(q^2) \gamma^\mu - f^A_2(q^2) i\sigma^{\mu q}/M_1
     + f^A_3(q^2) q^\mu/M_1 \Big]\gamma^5 u_1(p_1)\,, 
\label{eq:ff_def}
\ena 
where $q = p_1 - p_2$. Since we will also discuss lepton mass effects it is
necessary to also include the scalar form factors $f^V_3$ and $f^A_3$ in the
expansion~(\ref{eq:ff_def}).
The details of how to calculate the six form factors  
in the covariant confined quark model approach was discussed  
in our previous paper~\cite{Gutsche:2013pp,Gutsche:2013oea}. 

It is convenient to analyze the semileptonic decays of heavy baryons in terms 
of helicity amplitudes $H_{\lambda_2\lambda_j}$ which are linearly
related to the invariant form factors $f_i^V$ and $f_i^A$ (see details in 
Refs.~\cite{Kadeer:2005aq,Faessler:2009xn,Branz:2010pq,Gutsche:2013pp,%
Gutsche:2013oea}). 
Here we shall employ a generic notation such that the parent and daughter
baryons are denoted by $B_{1}$ and $B_{2}$.
The helicities of the daughter baryon $B_{2}$ and the effective current 
are denoted by $\lambda_2$ and $\lambda_j$, respectively. 
The pertinent relation is 
\be
H_{\lambda_2\lambda_j} =
\la B_2(\lambda_2) | \bar q^{\,\prime} \ O_\mu \, q | B_1(\lambda_1) \ra \, 
\epsilon^{\dagger\,\mu}(\lambda_j)
= H^{V}_{\lambda_2\lambda_j} - H^{A}_{\lambda_2\lambda_j}\,.  
\label{eq:hel_def}
\en 
The helicity amplitudes have been split into their  
vector $(H^V_{\lambda_2\lambda_j})$ 
and axial--vector $(H^A_{\lambda_2\lambda_j})$ parts. 
We shall work in the rest frame of 
the parent baryon $B_1$ with the daughter baryon $B_2$ moving in the
negative $z$-direction such that
$p_1^\mu = (M_1, {\bf 0})$, $p_2^\mu = (E_2, 0, 0,- |{\bf p}_2|)$ and
$q^\mu = (q_0, 0, 0,  |{\bf p}_2|)$. 
Further we use the following definitions of kinematical variables 
$q_0 = (M_+ M_- + q^2)/(2 M_1)$, 
$|{\bf p}_2| = \sqrt{Q_+Q_-}/{2 M_1}$  
and 
$E_2 = M_1 - q_0 = (M_1^2 + M_2^2 - q^2)/(2 M_1)$, where $q^2$ 
is the momentum squared transfered to the leptonic pair. 
We have introduced the notation 
$M_\pm = M_1 \pm M_2$, $Q_\pm = M_\pm^2 - q^2$. 
Angular momentum conservation fixes the helicity $\lambda_1$
of the parent baryon such that $\lambda_1 = - \lambda_2 + \lambda_j$. 
The relations between the helicity amplitudes $H^{V,A}_{\lambda_2\lambda_j}$ 
and the invariant amplitudes are given 
by~\cite{Gutsche:2013pp,Gutsche:2013oea} 
\bea 
H^V_{\pm\frac{1}{2} \pm 1} &=& \sqrt{2 Q_-} \, 
\biggl( f_1^V + \frac{M_+}{M_1} \, f_2^V \biggr)\,, 
\qquad\hspace*{.68cm} 
H^A_{\pm\frac{1}{2} \pm 1} \ = \ \pm \sqrt{2 Q_+} \, 
\biggl( f_1^A - \frac{M_-}{M_1} \, f_2^A \biggr)\,,
\nn
H^V_{\pm\frac{1}{2}  0} &=& \sqrt{\frac{Q_-}{q^2}} \,  
\biggl( M_+ \, f_1^V + \frac{q^2}{M_1} \, f_2^V \biggr)\,, 
\qquad\hspace*{.45cm} 
H^A_{\pm\frac{1}{2} 0} \ = \ \pm\sqrt{\frac{Q_+}{q^2}} \,  
\biggl( M_- \, f_1^A  - \frac{q^2}{M_1} \, f_2^A \biggr)\,, 
\label{eq:hel_ff} \\
H^V_{\pm\frac{1}{2} \pm t} &=& \pm \sqrt{\frac{Q_+}{q^2}} \, 
\biggl( M_- \, f_1^V + \frac{q^2}{M_1} \, f_3^V \biggr)\,, 
\qquad
H^A_{\pm\frac{1}{2} \pm t} \ = \ \pm \sqrt{\frac{Q_-}{q^2}} \, 
\biggl( M_+ \, f_1^A - \frac{q^2}{M_1} \, f_3^A \biggr)\,. 
\nonumber
\ena 
The scalar helicity component is denoted by $\lambda_{j}=t$ . The scalar 
helicity amplitudes contribute only for nonzero charged lepton masses.
As in Ref.~\cite{Gutsche:2013pp} we  introduce the following 
combinations of helicity amplitudes 
\bea 
\begin{array}{lr}
\mbox{$ H_U = |H_{\frac{1}{2}1}|^2 +  |H_{-\frac{1}{2}-1}|^2$} &
\hfill\mbox{ \rm transverse unpolarized}\quad(\mathrm{pc})\,, 
\\
\mbox{$ H_L = |H_{\frac{1}{2}0}|^2 +  |H_{-\frac{1}{2}0}|^2$} &
\hfill\mbox{ \rm longitudinal unpolarized}\quad(\mathrm{pc})\,,  
\\
\mbox{$ H_S = |H_{\frac{1}{2}t}|^2 +  |H_{-\frac{1}{2}t}|^2$} &
\hfill\mbox{ \rm scalar unpolarized}\quad(\mathrm{pc})\,. 
\\
\mbox{$ H_P = |H_{\frac{1}{2}1}|^2 -  |H_{-\frac{1}{2}-1}|^2$} &
\hfill\mbox{ \rm transverse parity--odd polarized}\quad(\mathrm{pv})\,, 
\\
\mbox{$ H_{L_P} = |H_{\frac{1}{2}0}|^2 -  |H_{-\frac{1}{2}0}|^2$} &
\hfill\mbox{ \rm longitudinal polarized}\quad(\mathrm{pv})\,,  
\\
\mbox{$ H_{S_P} = |H_{\frac{1}{2}t}|^2 -  |H_{-\frac{1}{2}t}|^2$} &
\hfill\mbox{ \rm scalar polarized}\quad(\mathrm{pv})\,, 
\\
\mbox{$ H_{LS} = H_{\frac{1}{2}t} H_{\frac{1}{2}0} + 
H_{-\frac{1}{2}t} H_{-\frac{1}{2}0}$} &
\hfill\mbox{ \rm longitudinal-scalar interference}\quad(\mathrm{pc})\,. 
\\
\end{array}
\label{eq:hel_comb}
\ena   
We have indicated the parity properties of the seven combinations in round 
brackets.
The partial helicity width $\Gamma_{I}$ and branching ratio $B_{I}$  
corresponding to one of the seven specific combinations
of differential helicity amplitudes in~(\ref{eq:hel_comb}) are defined as 
\bea
\Gamma_I &=&
\int\limits_{m_\ell^2}^{M^2_-} dq^2 \
\frac{d\Gamma_I}{dq^2} \,, \quad \quad 
B_I = \Gamma_I \ \tau\,, \nonumber\\
\frac{d\Gamma_I}{dq^2}
&=& \frac{1}{2} \, \frac{G_F^2}{(2 \pi)^3} \, |V_{Qq}|^2 \,
\frac{|{\bf p}_2|}{12 M_1^2} \, q^2 \,
\biggl(1 - \frac{m_\ell^2}{q^2}\biggr)^2 \, H_I\,, 
\quad\quad\quad I = U, L, S, P, L_P, S_P, LS \,,
\label{eq:hel_partial_width}
\ena 
where $\tau$ is the lifetime of the parent baryon: 
$\tau_{\Lambda_b} = 1.425 \times 10^{-12}$ s and 
$\tau_{\Lambda_c} = 0.2   \times 10^{-12}$ s. 
For the $\Lambda_b \to p + \ell^- \bar\nu_\ell$ 
and $\Lambda_c \to n + \ell^+ \nu_\ell$ decay widths and 
the asymmetry parameter $\alpha_{F\!B}^\ell$ 
(forward-backward asymmetry of the charged leptons in 
the $W^-{\rm off-shell}$ rest frame or in the $(\ell, \nu_\ell)$ c.m. 
frame) one finds~\cite{Kadeer:2005aq} 
\bea
\Gamma  &=& \int\limits_{m_\ell^2}^{M^2_-} dq^2 \
\frac{d\Gamma}{dq^2} \,, \nonumber\\
d\Gamma &=& d\Gamma_U
+ d\Gamma_L + \frac{m_\ell^2}{2q^2}
\biggl( d\Gamma_U + d\Gamma_L + 3d\Gamma_S\biggr)
\label{eq:LQN_decay}
\ena
and 
\bea 
\alpha_{F\!B}^\ell &=& \frac{\tilde\Gamma}{\Gamma} \,, \quad 
\tilde\Gamma  \ = \ \int\limits_{m_\ell^2}^{M^2_-} dq^2 \ 
\frac{d\tilde\Gamma}{dq^2} \,, \nonumber\\
d\tilde\Gamma &=& 
\frac{3}{4} \biggl\{\pm d\Gamma_P - 
\frac{2m_\ell^2}{q^2} d\Gamma_{LS} 
\biggr\} \,. 
\label{eq:alpha_as}
\ena 
where the plus/minus signs refers to the 
$\Lambda_b \to p \ell^- \bar\nu_\ell$ and 
$\Lambda_c \to n \ell^+ \nu_\ell$ cases, 
respectively~\cite{Kadeer:2005aq}. 

\section{The \boldmath{$\Lambda_Q\to N$} transitions in the
covariant confined quark model} 
\label{sec:form_factors}

For the description of the couplings of the heavy baryons 
$\Lambda_Q$ ($Q=b,c$) and nucleons to their constituent quarks 
we employ generic Lagrangians 
which~\cite{Ivanov:1996fj,Ivanov:1996pz,Gutsche:2012ze} 
\bea
\Lambda_Q:\qquad &&{\cal L}^{\Lambda_Q}_{\rm int}(x) 
 = g_{\Lambda_Q} \,\bar \Lambda_Q(x)\cdot J_{\Lambda_Q}(x) + \mathrm{H.c.}\,, 
\label{eq:lag_Lambda}\\
\phantom{\Lambda_Q:}\qquad &&J_{\Lambda_Q}(x) 
= \int\!\! dx_1 \!\! \int\!\! dx_2 \!\! \int\!\! dx_3 \, 
F_{\Lambda_Q}(x;x_1,x_2,x_3) \, 
\epsilon^{a_1a_2a_3} \, Q^{a_1}(x_1)\,u^{a_2}(x_2) \,C \, \gamma^5 \, 
d^{a_3}(x_3)\,,
\nn
\nn
N:\qquad &&{\cal L}^{N}_{\rm int}(x) 
 = g_{N} \,\bar N(x)\cdot J_{N}(x) + \mathrm{H.c.}\,, 
\label{eq:lag_N}\\
\phantom{N:}\qquad &&
J_N(x) = (1-x_N) J_N^V + x_N J_N^T \,, \nonumber\\
\phantom{N:}\qquad && 
J_{p}^V(x)
= \int\!\! dx_1 \!\! \int\!\! dx_2 \!\! \int\!\! dx_3 \, 
F_{N}(x;x_1,x_2,x_3) \, 
\epsilon^{a_1a_2a_3} \, \gamma^\mu \gamma^5 \, 
d^{a_1}(x_1)\,u^{a_2}(x_2) \,C \, \gamma_\mu \, 
u^{a_3}(x_3)\,,\nonumber\\
\phantom{N:}\qquad &&
J_{p}^T(x)
= \int\!\! dx_1 \!\! \int\!\! dx_2 \!\! \int\!\! dx_3 \, 
F_{N}(x;x_1,x_2,x_3) \, 
\epsilon^{a_1a_2a_3} \, \frac{1}{2} \, \sigma^{\mu\nu} \gamma^5 \, 
d^{a_1}(x_1)\,u^{a_2}(x_2) \,C \, \sigma_{\mu\nu} \, 
u^{a_3}(x_3)\,,\nonumber\\
\phantom{N:}\qquad && 
J_{n}^V(x)
= - \int\!\! dx_1 \!\! \int\!\! dx_2 \!\! \int\!\! dx_3 \, 
F_{N}(x;x_1,x_2,x_3) \, 
\epsilon^{a_1a_2a_3} \, \gamma^\mu \gamma^5 \, 
u^{a_1}(x_1)\,d^{a_2}(x_2) \,C \, \gamma_\mu \, 
d^{a_3}(x_3)\,,\nonumber\\
\phantom{N:}\qquad &&
J_{n}^T(x)
= - \int\!\! dx_1 \!\! \int\!\! dx_2 \!\! \int\!\! dx_3 \, 
F_{N}(x;x_1,x_2,x_3) \, 
\epsilon^{a_1a_2a_3} \, \frac{1}{2} \, \sigma^{\mu\nu} \gamma^5 \, 
u^{a_1}(x_1)\,d^{a_2}(x_2) \,C \, \sigma_{\mu\nu} \, 
d^{a_3}(x_3)\,. 
\nonumber 
\ena
The color index is denoted by $a$ and 
$C = \gamma^0\gamma^2$ is the charge 
conjugation matrix. In the $\Lambda_Q$ baryon case we take 
the $u$ and $d$ quarks to be in a $S=0$ and $I=0$ $[ud]$ diquark 
configuration antisymmetric in spin and isospin. In case of the nucleon 
we use an interpolating current of the the so-called {\it vector} 
current variety~\cite{Gutsche:2012ze}, 
which contains two $u$ quarks (in case of the proton) or  
two $d$ quarks (in case of the neutron) in a symmetric 
diquark configuration with spin and isospin equal to 1. 

The nonlocal vertex functions in momentum space are denoted by 
$\bar\Phi_H(-P^2)$ and  
are obtained from the Fourier transformations of the vertex functions 
$F_H$ entering in Eqs.~(\ref{eq:lag_Lambda}) and (\ref{eq:lag_N}). 
In the numerical calculations we choose a simple Gaussian form 
for the vertex functions (for both mesons and baryons): 
\be
\bar\Phi_H(-P^2) = \exp(P^{\,2}/\Lambda_H^2) \,,
\label{eq:Gauss}
\en  
where $\Lambda_H$ is a size parameter describing the size of the distribution 
of the quarks inside a given hadron $H$. 
The values for these parameters were fixed before 
in~\cite{Ivanov:2011aa,Gutsche:2012ze,Dubnicka:2013vm,%
Gutsche:2013pp,Gutsche:2013oea}. We would like to stress 
that the Minkowskian momentum variable $P^{\,2}$ turns into the Euclidean form 
$-\,P^{\,2}_E$ needed for the the appropriate fall--off behavior of the 
correlation function~(\ref{eq:Gauss}) in the Euclidean region.
We emphasize that any choice for the correlation function $\bar\Phi_H$ is 
acceptable
as long as it falls off sufficiently fast in the ultraviolet region of
Euclidean space. The choice of a Gaussian form for $\bar\Phi_H$ has obvious
calculational advantages.

\begin{figure}
\begin{center}
\epsfig{figure=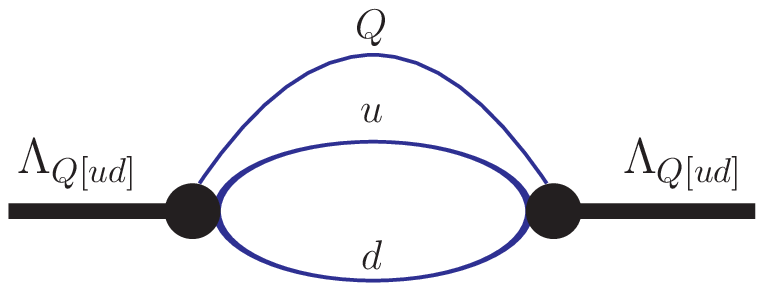,scale=.8}
\caption{Mass operator of $\Lambda_{Q}$ baryon.} 
\label{fig:fig_mass}

\epsfig{figure=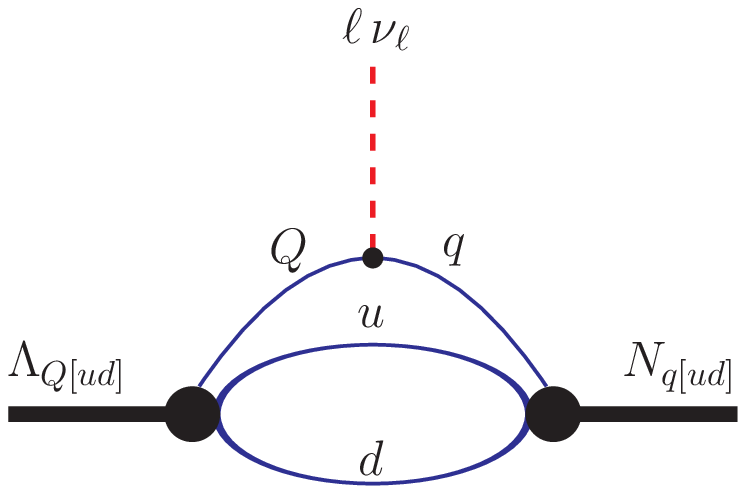,scale=.8} 
\caption{Diagrams contributing to the semileptonic 
decays of $\Lambda_Q$ baryons.}  
\label{fig:fig_sl}
\end{center}
\end{figure}

For given values of the size parameters $\Lambda_H$
the coupling constants $g_{\Lambda_{Q}}$ and $g_{N}$ are determined by 
the compositeness condition suggested by Weinberg~\cite{Weinberg:1962hj}
and Salam~\cite{Salam:1962ap} (for a review, see~\cite{Hayashi:1967hk})
and extensively used in our approach (for details, see~\cite{Efimov:1993ei}).  
The compositeness condition
implies that the renormalization constant of 
the hadron wave function is set equal to zero: 
\be 
Z_H = 1 - \Sigma^\prime_H  = 0 \,,  
\label{eq:Z=0}
\en 
where $\Sigma^\prime_H$ is the on-shell derivative of the 
hadron mass function $\Sigma_H$ with respect to its momentum. 
In Fig.~\ref{fig:fig_sl} we present as an example the diagram 
corresponding to the mass operator of the $\Lambda_Q$ baryon.  
The compositeness condition can be seen to provide for the correct charge 
normalization for a charged bound state (see e.g.\cite{Ivanov:2011aa}).

How to calculate the matrix element of the baryonic 
transitions has been discussed in detail in our 
previous papers~\cite{Gutsche:2012ze,Gutsche:2013pp,Gutsche:2013oea}. 
In our approach semileptonic transitions between baryons 
are described by a two-loop Feynman-type
diagram involving nonlocal vertex functions as shown in Fig.~\ref{fig:fig_sl}. 

In the calculation of quark-loop diagram (Fig.~\ref{fig:fig_sl}) 
we use the set of model parameters fixed in our previous studies.  
The model parameters are the constituent quark masses $m_q$ and
the infrared cutoff parameter $\lambda$ responsible for quark confinement.
They are taken from a fit done in 
the papers~\cite{Ivanov:2011aa,Gutsche:2012ze,Dubnicka:2013vm}:
\be
\def\arraystretch{2}
\begin{array}{ccccccc}
     m_u        &      m_s        &      m_c       &     m_b & \lambda  &
\\\hline
 \ \ 0.235\ \   &  \ \ 0.424\ \   &  \ \ 2.16\ \   &  \ \ 5.09\ \   &
\ \ 0.181\ \   & \ {\rm GeV}
\end{array}
\label{eq: fitmas}
\en
The dimensional size parameters $\Lambda$ in Eq.~(\ref{eq:Gauss})
and the dimensionless parameter $x_N$ in Eq.~(\ref{eq:lag_N}) 
characterizing the vector and tensor current mixing
have been determined in~\cite{Gutsche:2012ze,Gutsche:2013pp} 
by a fit to the magnetic moments of nucleons and to the semileptonic 
decays $\Lambda_{b} \to \Lambda_{c}  \ell^{-} \bar \nu_{\ell}$ and 
$\Lambda_{c} \to \Lambda  \ell^{+} \nu_{\ell}$. The resulting values are

\be
\def\arraystretch{2}
\begin{array}{c|ccccc}
x_N & \Lambda_N &  \Lambda_{\Lambda_s} 
& \Lambda_{\Lambda_c} & \Lambda_{\Lambda_b} &
\\\hline
 \ \ 0.8\ \   &  \ \ 0.50  \ \   &  \ \ 0.492 \ \   &  
\ \ 0.867 \ \ & \ \ 0.571 & \ {\rm GeV}  
\end{array}
\label{eq:fitlam}
\en

It should be clear that the evaluation of the form factors is 
technically quite intricate. 
It involves the calculation of a two-loop Feynman diagram
with a complex spin structure resulting from the quark propagators and the
vertex functions which leads to a number of two-loop tensor 
integrals. In order to tackle this difficult
task we have automated the calculation in the form  
of FORM and FORTRAN packages written for this purpose.

\begin{figure}[htb]
\begin{center}
\epsfig{figure=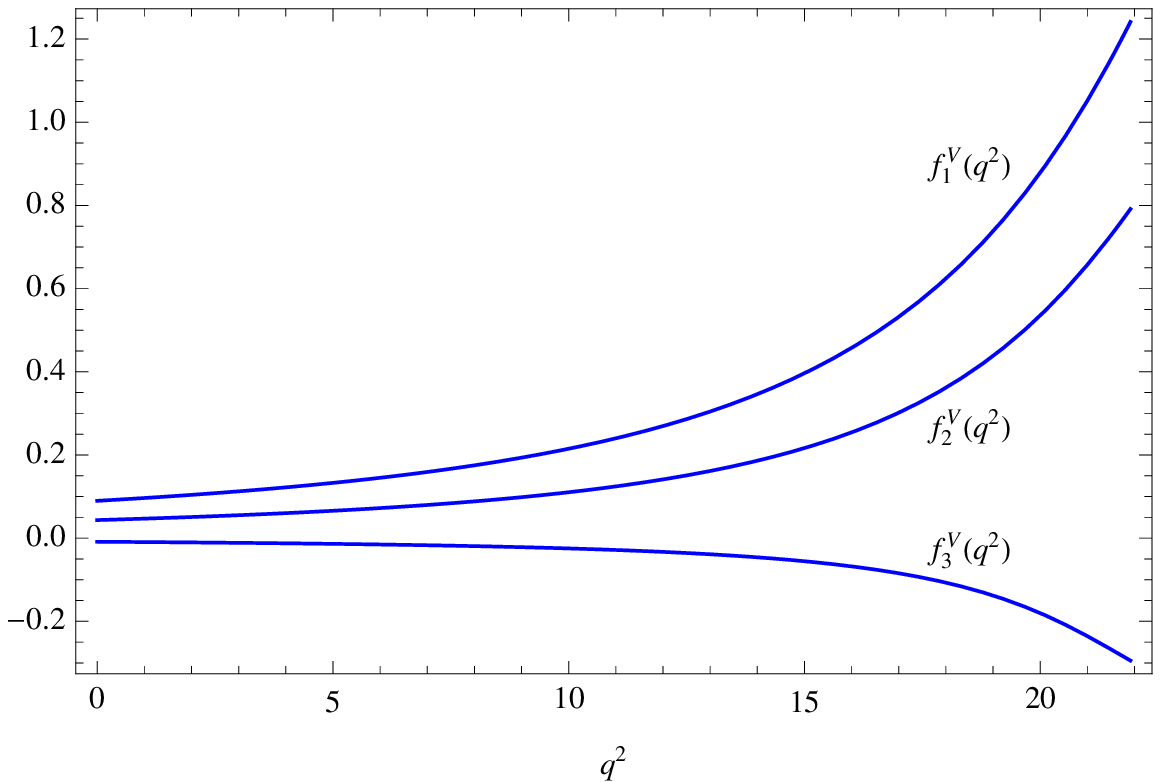,scale=.65}
\caption{$q^{2}$-dependence of the vector form factors 
for the $\Lambda_b \to p$ transition} 
\label{fig:VVbu}

\vspace*{.2cm}
\epsfig{figure=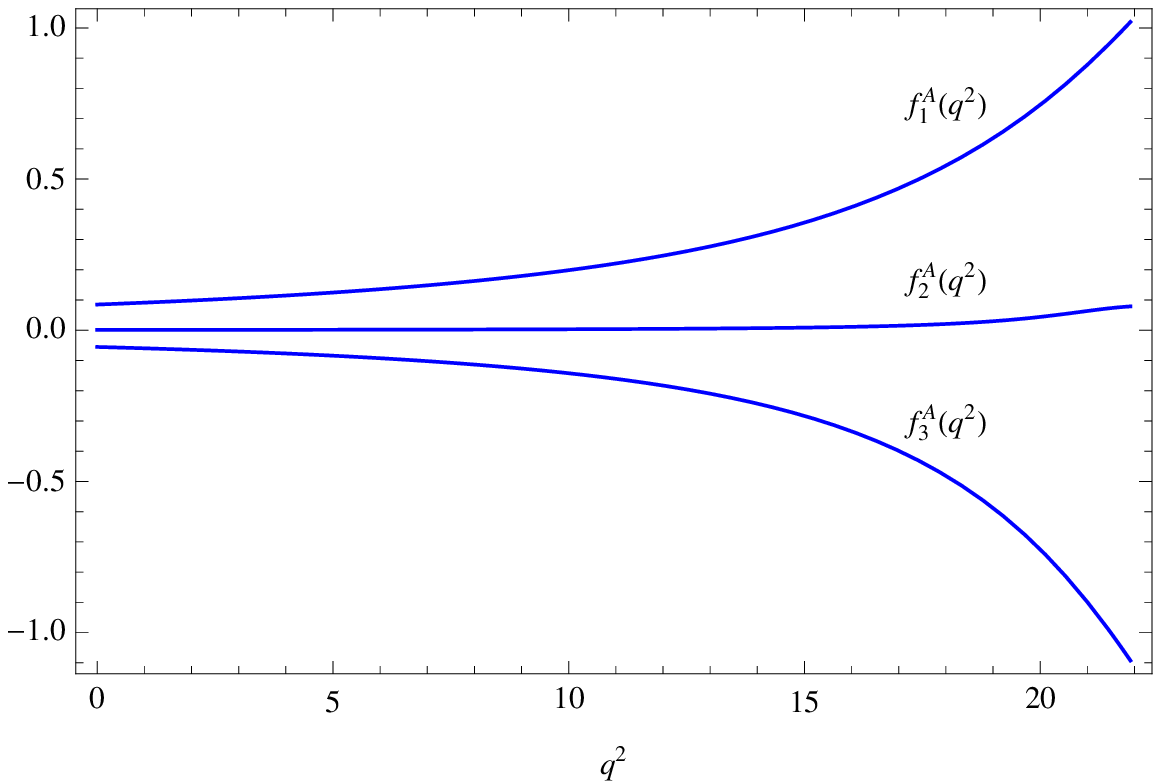,scale=.65}
\caption{$q^{2}$-dependence of the axial form factors 
for the $\Lambda_b \to p$ transition} 
\label{fig:AAbu}
\end{center}
\end{figure}

\begin{figure}
\begin{center}
\epsfig{figure=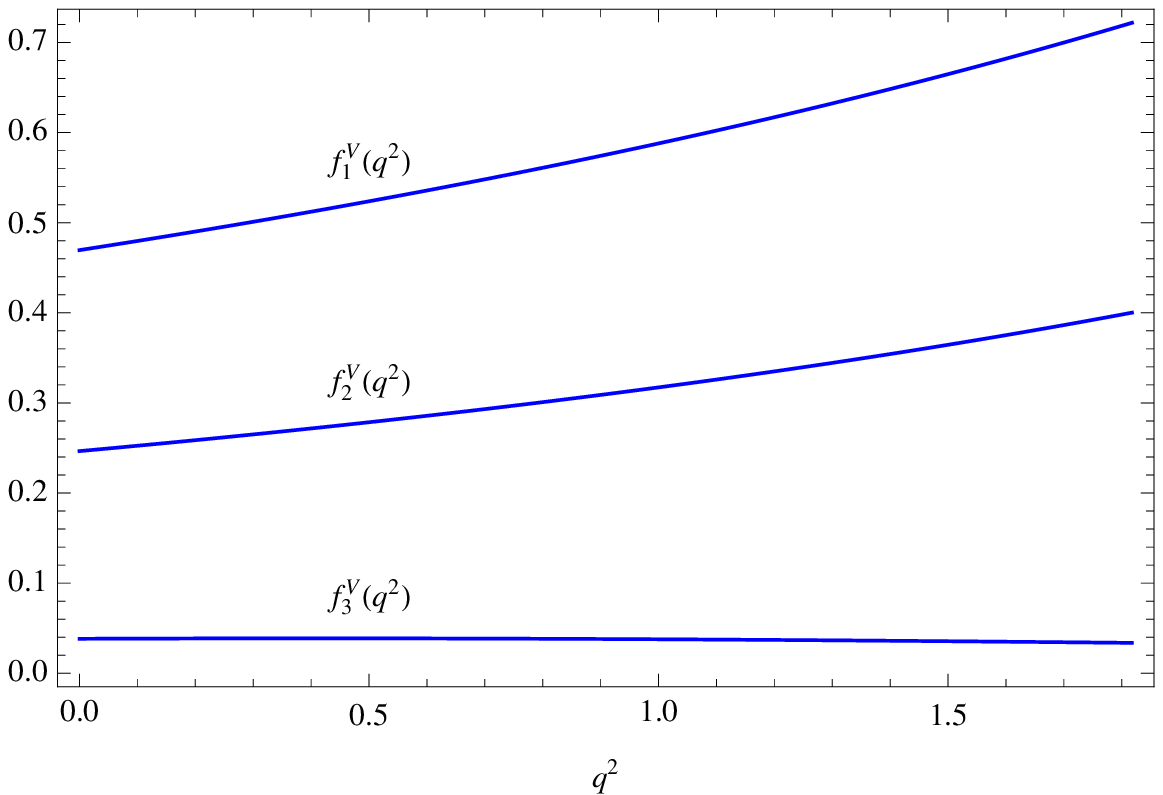,scale=.65}
\caption{$q^{2}$-dependence of the vector form factors 
for the $\Lambda_c \to n$ transition} 
\label{fig:VVcd}

\vspace*{.2cm}

\epsfig{figure=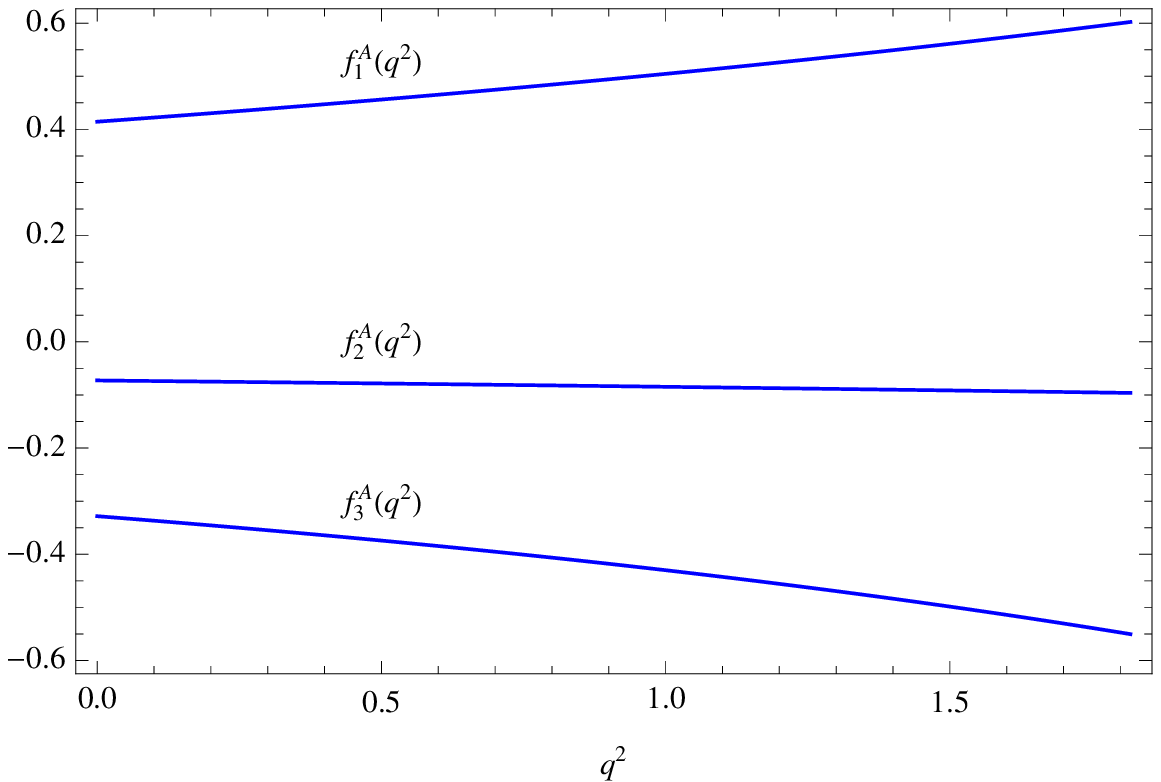,scale=.65}
\caption{$q^{2}$-dependence of the axial form factors 
for the $\Lambda_c \to n$ transition} 
\label{fig:AAcd}
\end{center}
\end{figure}

The $q^{2}$--behavior of the form factors are shown in 
Figs.~(\ref{fig:VVbu})-(\ref{fig:AAcd}).
The results of our numerical calculations are well represented
by a double--pole parametrization of the form 
\be\label{DPP} 
f(\hat{s})= f(0) \, \frac{1}{1 - a \hat{s} + b \hat{s}^2}\,, 
\en
where $\hat{s} = q^2/M_1^2$. 
Using such a parametrization accelerates the necessary 
$q^{2}$--integrations which can be done using the parametrization~(\ref{DPP})
without having to do a numerical evaluation of the loop diagram for each 
$q^{2}$ value separately.
The values of $f(0)$, $a$ and $b$ are listed  
in TABLES~\ref{tab:fflbu} and~\ref{tab:fflcd}. 
Note that the dominant form factors $f_{1}^{V/A}$ in TABLES~\ref{tab:fflbu} 
and~\ref{tab:fflcd} are very close to a dipole form since one has
$\sqrt{b}\sim a/2$ in all four cases. The effective dipole mass is given by
$m_{eff}=M_{1}/\sqrt{a/2}$ or $m_{eff}=M_{1}/b^{1/4}$. In the 
$\Lambda_{b}\to p$ case the effective dipole mass is very close to the 
average of
the $B,B^{\ast}$ meson masses. In the $\Lambda_{c} \to n$ case the 
effective dipole mass is about $50\,\%$ higher than the average of
the $D,D^{\ast}$ meson masses. 

\newpage 

\begin{table}
\caption{Parameters for the approximated form factors
in Eqs.~(\ref{DPP}) for  the $\Lambda_b \to p$ transitions.} 
\begin{center}
\def\arraystretch{1.2}
\begin{tabular}{ccccccc}
\hline
       &\qquad $f_1^V$ \qquad 
       &\qquad $f_2^V$ \qquad 
       &\qquad $f_3^V$ \qquad 
       &\qquad $f_1^A$ \qquad 
       &\qquad $f_2^A$ \qquad 
       &\qquad $f_3^A$ \qquad \\
\hline
$f(0)$           & 0.090 & 0.043 & $-$ 0.009 & 0.085 & 0.001 & $-$ 0.055 \\
$a$              & 2.262 & 2.380 & 2.592 & 2.213 & 2.793  & 2.403  \\
$b$              & 1.333 & 1.466 & 1.720 & 1.286 & 1.976  & 1.491  \\
\hline
\end{tabular}
\label{tab:fflbu} 
\end{center}

\vspace*{.2cm}

\caption{Parameters for the approximated form factors
in Eq.~(\ref{DPP}) for the $\Lambda_c \to n$ transitions.} 
\begin{center}
\def\arraystretch{1.2}
\begin{tabular}{ccccccc}
\hline
       &\qquad $f_1^V$ \qquad 
       &\qquad $f_2^V$ \qquad 
       &\qquad $f_3^V$ \qquad 
       &\qquad $f_1^A$ \qquad 
       &\qquad $f_2^A$ \qquad 
       &\qquad $f_3^A$ \qquad \\
\hline
$f(0)$           & 0.470 & 0.247 & 0.038 & 0.414 & $-$ 0.073 & $-$ 0.328 \\
$a$              & 1.111 & 1.240 & 0.308 & 0.978 & 0.781  & 1.330  \\
$b$              & 0.303 & 0.390 & 1.998 & 0.235 & 0.225  & 0.486  \\
\hline
\end{tabular}
\label{tab:fflcd}
\end{center}

\vspace*{.2cm}

\caption{$\Lambda_b \to p$ transitions: 
Comparison of our form factors at $q^{2}=0$ and $q^2 = q^2_{\rm max}$ 
with those obtained in \cite{Wei:2009np,Khodjamirian:2011jp}.  
}
\begin{center}
\def\arraystretch{1.2}
\begin{tabular}{cccccccc}
\hline
  &  &  \qquad $f_1^V$ \qquad
     &  \qquad $f_2^V$ \qquad
     &  \qquad $f_3^V$ \qquad
     &  \qquad $f_1^A$ \qquad
     &  \qquad $f_2^A$ \qquad
     &  \qquad $f_3^A$ \qquad
\\
\hline
$q^2=0$  & \cite{Khodjamirian:2011jp}  \qquad &
         \quad   $0.12^{+0.03}_{-0.04}$ & 
         $0.047^{+0.015}_{-0.013}$ & $-$ & 
         $0.12^{+0.03}_{-0.03}$ & 
         $-\,0.016^{+0.007}_{-0.005}$ & 
         $-$ \\
        & \cite{Wei:2009np}  \qquad &
         \quad   0.1131 & 0.0356 & $-$ & 0.1112 &  $-$ 0.0097 & $-$ \\
        & our  \qquad& 0.080  & 0.036  & $-$ 0.005 & 0.077 & $-$ 0.001 
        & $-$ 0.046 \\
\hline
$q^2=q^2_{\rm max}$  & \cite{Wei:2009np}  \qquad &
         \quad   0.626 & 0.231 & $-$ & 0.581 &  $-$ 0.089 & $-$ \\
    & our \qquad & 1.254 & 0.801 & $-$ 0.300 & 1.030 & 0.082 & $-$ 1.105 \\
\hline
\end{tabular}
\label{tab:MR_bu}
\end{center}

\vspace*{.2cm} 

\caption{$\Lambda_c \to n$ transitions: 
Comparison of our form factors values  at
$q^{2}=0$ and $q^2 = q^2_{\rm max}$ to 
those obtained in~\cite{Wei:2009np,Khodjamirian:2011jp}. 
}
\begin{center}
\def\arraystretch{1.2}
\begin{tabular}{cccccccc}
\hline
  &  &  \qquad $f_1^V$ \qquad
     &  \qquad $f_2^V$ \qquad
     &  \qquad $f_3^V$ \qquad
     &  \qquad $f_1^A$ \qquad
     &  \qquad $f_2^A$ \qquad
     &  \qquad $f_3^A$ \qquad
\\
\hline
$q^2=0$  & \cite{Khodjamirian:2011jp}  \qquad &
         \quad   $0.59^{+0.15}_{-0.11}$ & 
         $0.43^{+0.13}_{-0.12}$ & $-$ & 
         $0.55^{+0.14}_{-0.15}$ & 
         $-\,0.16^{+0.08}_{-0.05}$ & 
         $-$ \\
        & \cite{Wei:2009np}  \qquad &
         \quad   0.1081   & 0.0311 & $-$ & 0.1065 &  $-$ 0.0064 & $-$ \\
    & our  \quad & 0.470  & 0.246  & 0.039 & 0.414 & $-$ 0.073
& $-$ 0.328 \\
\hline
$q^2=q^2_{\rm max}$ & \cite{Wei:2009np}  \qquad &
         \quad 0.187 & 0.0652 & $-$ & 0.187 &  $-$ 0.0214 &  $-$ \\ 
    & our \qquad & \quad 0.721 & 0.400 &  0.033 & 0.602 & $-$ 0.096
    & $-$ 0.550\\
\hline
\end{tabular}
\label{tab:MR_cd}
\end{center}
\end{table} 

\begin{figure}
\begin{center}
\epsfig{figure=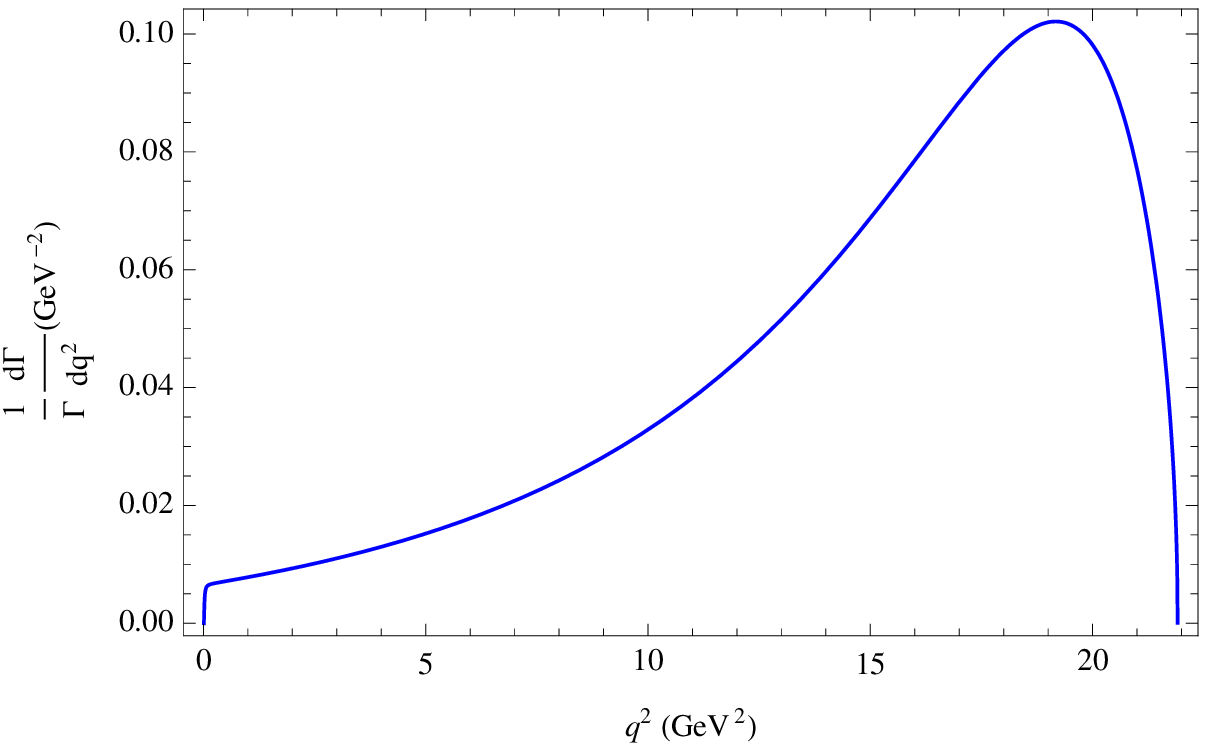,scale=.8}
\caption{$q^{2}$-distribution for the 
$\Lambda_b^0 \to p \mu^- \bar\nu_\mu$ transition} 
\label{fig:dGbu}

\vspace*{.5cm}

\epsfig{figure=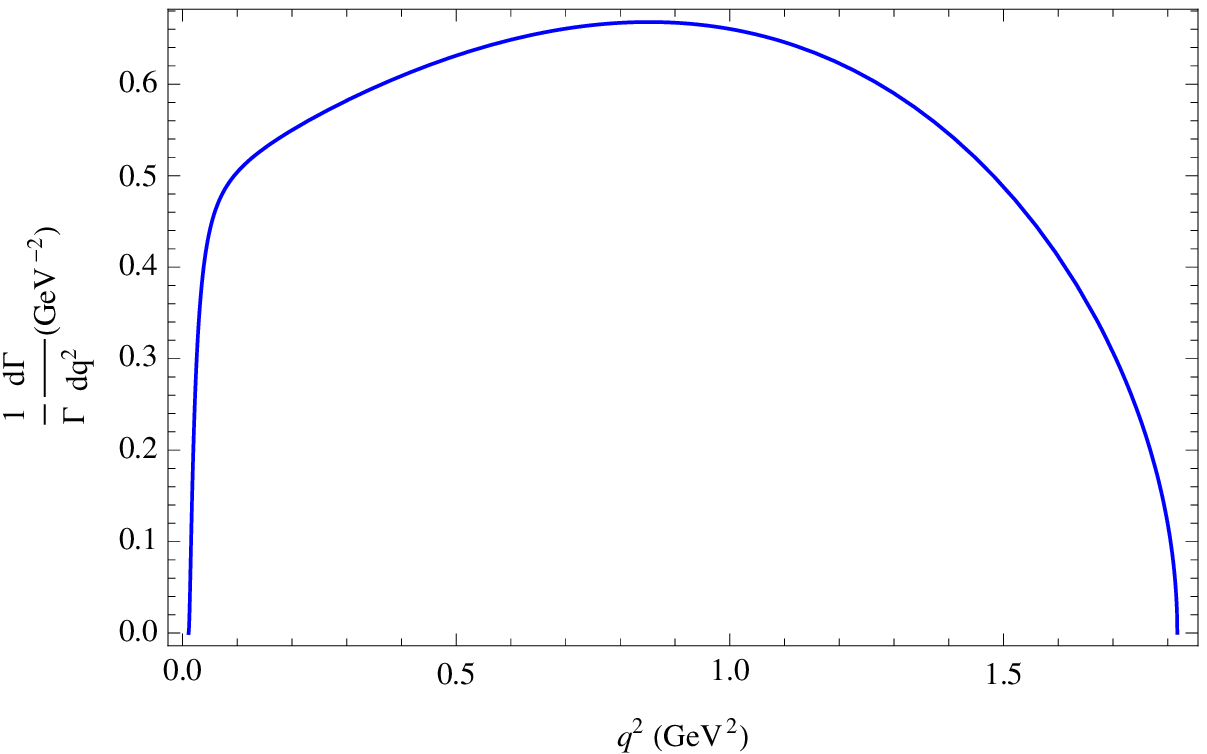,scale=.8}
\caption{$q^{2}$-distribution for the 
$\Lambda_c^+ \to n \mu^+ \nu_\mu$ transition} 
\label{fig:dGcd}
\end{center}
\end{figure}

In TABLES~\ref{tab:MR_bu} and \ref{tab:MR_cd} we list our form factor 
results for $q^{2} = 0$ and $q^2 = q^2_{\rm max}$ 
and compare them to the results of the light-front
diquark model calculation of~\cite{Wei:2009np} and 
the QCD light-cone sum rules of~\cite{Khodjamirian:2011jp}. 

It is interesting to explore how the present form factors are related to
the corresponding charged or neutral current form factors 
$\Lambda_{Q}\to\Lambda$.
In the limit of $SU(3)$ the $\Lambda_{Q}\to\Lambda$ and
$\Lambda_{Q}\to N$ ($N = p,n$) form factors are related by 
$F(\Lambda_{Q}\to\Lambda)=\sqrt{2/3} \,F(\Lambda_{Q}\to N)$. This can be 
seen by using the $\bar 3 \otimes 3\,\to 8$ Clebsch-Gordan table listed in 
\cite{Kaeding:1995vq}. Based on the observation that the $[ud]$ diquark
is the $(Y=2/3,I=0)$ member of the $\bar 3$ multiplet one needs the
C.G. coefficients 
\begin{eqnarray}
\label{clebsch}
\mbox{$\Lambda_{Q}\to\Lambda$}&:&
\qquad <\mbox{\boldmath $\bar 3$},-\tfrac 2 3,0,0;\,\mbox{\boldmath $3$} ,
\tfrac 2 3,0,0|\,\mbox{\boldmath $8$},0,0,0>\,=\sqrt{2/3}\, , \\
\mbox{$\Lambda_{Q}\to N$}&:&\qquad <\mbox{\boldmath $\bar 3$},
\tfrac 2 3,0,0;\, \mbox{\boldmath $3$},\tfrac 1 3,
\tfrac 1 2,\tfrac 1 2|\,
\mbox{\boldmath $8$},1,\tfrac 1 2,\tfrac 1 2>\,\,\,=1\,.
\nonumber
\end{eqnarray}
The labelling in~(\ref{clebsch}) proceeds according to the sequence 
$|\,\mbox{\boldmath $R$},Y,I,I_{z}>$
where $\mbox{\boldmath $R$}$ denotes the relevant $SU(3)$ representation.
As a check on our calculations we have obtained the same result analytically
in the $SU(3)$ limit by setting 
\begin{eqnarray} 
\Lambda_{\Lambda_s} = \Lambda_N\,, \quad M_{\Lambda} = M_N\,, \quad 
m_s = m_u\,. 
\end{eqnarray}  

Our predictions for the branching ratios of the heavy-to-light 
transitions are listed in TABLE~\ref{tab:rates_our}. 
In TABLES~\ref{tab:rates_bu} and \ref{tab:rates_cd} we compare 
our results for the rates (in units of ($|V_{qq^\prime}|^2$ \, ps$^{-1}$)) 
with the predictions of other theoretical approaches. We use the 
compilations of results given in Ref.~\cite{Azizi:2009wn}. 
The results for the lepton-side asymmetry parameters $\alpha_{F\!B}^\ell$ 
are shown in TABLE~\ref{tab:asym}. 

\begin{table}
\begin{center}
\caption{Total $B(\Lambda_Q \to N \ell \nu_\ell)$ 
and partial $B_I(\Lambda_Q \to N \ell \nu_\ell)$ 
helicity contributions to branching ratios (in units of $10^{-4}$).} 

\vspace*{.15cm}

\def\arraystretch{1.25}
    \begin{tabular}{l|c|c|c|c}
      \hline
& \multicolumn{4}{c}{Our results} \\
\cline{2-5}
\qquad Mode \qquad & \qquad $B$ \qquad  
                   & \qquad $B_U$ \qquad  
                   & \qquad $B_L$ \qquad  
                   & \qquad $B_S$ \qquad  \\ 
\hline
$\Lambda_b \to p e^- \bar\nu_e$ \qquad 
& 2.9 & 1.6 & 1.3 & $\simeq 0$  \\
\hline 
$\Lambda_b \to p \mu^- \bar\nu_\mu$ \qquad 
& 2.9 & 1.6 & 1.3 & $\simeq 0$  \\
\hline
$\Lambda_b \to p \tau^- \bar\nu_\tau$ \qquad 
& 2.1 & 1.0 & 0.7 & 0.2  \\
\hline 
$\Lambda_c \to n e^+ \nu_e$ \qquad 
& 20.7 & 8.3  & 12.3 & $\simeq 0$  \\
\hline
$\Lambda_c \to n \mu^+ \nu_\mu$ 
& 20.2 & 8.1  & 11.3 & 0.5  \\
\hline
\end{tabular}
\label{tab:rates_our}
\end{center}

\vspace*{.25cm} 

\begin{center}
\caption{Decay widths $\Lambda_b \to p \ell^- \bar\nu_\ell$ 
(in units of $|V_{ub}|^2$ ps$^{-1}$).} 

\vspace*{.15cm}

\def\arraystretch{1}
    \begin{tabular}{c|c|c}
      \hline
Mode & Our result  & Theoretical predictions \\
\hline
$\Lambda_b \to p e^- \bar\nu_e$ 
    & 13.3 & 6.48~\cite{Datta:1995mv}; 7.47~\cite{Ivanov:1996fj}; 
      4.55~\cite{Pervin:2005ve};  
      7.55~\cite{Pervin:2005ve}; 11.8~\cite{Wei:2009np}; 
     $19.0^{+8.6}_{-6.9}$~\cite{Khodjamirian:2011jp};\\ 
    & & $250 \pm 85$~\cite{Azizi:2009wn}; 
        $235 \pm 85$~\cite{Azizi:2009wn}; 
        $477 \pm 175$~\cite{Azizi:2009wn}; 
        $376 \pm 125$~\cite{Azizi:2009wn}\\
\hline
$\Lambda_b \to p \mu^- \bar\nu_\mu$ 
    & 13.3 & 6.48~\cite{Datta:1995mv}; 7.47~\cite{Ivanov:1996fj}; 
      4.55~\cite{Pervin:2005ve};  
      7.55~\cite{Pervin:2005ve}; \\ 
    & &11.8~\cite{Wei:2009np}; 
    $19.0^{+8.6}_{-6.9}$~\cite{Khodjamirian:2011jp}; 
         20.5~\cite{Huang:1998rq}; 
         25.8~\cite{Marques de Carvalho:1999ia}; 
      36.5~\cite{Huang:2004vf}; 
      56.2~\cite{Huang:2004vf}; \\
    & &
     $250 \pm 85$~\cite{Azizi:2009wn}; 
     $235 \pm 85$~\cite{Azizi:2009wn}; 
     $478 \pm 175$~\cite{Azizi:2009wn}; 
     $384 \pm 125$~\cite{Azizi:2009wn} \\
\hline
$\Lambda_b \to p \tau^- \bar\nu_\tau$ 
      & 9.6 & 4.01~\cite{Pervin:2005ve};  
      6.55~\cite{Pervin:2005ve}; 
      $312 \pm 105$~\cite{Azizi:2009wn}; 
      $208 \pm 70$~\cite{Azizi:2009wn}; 
      $646 \pm 215$~\cite{Azizi:2009wn}; 
      $193 \pm 70$~\cite{Azizi:2009wn}\\
\hline
\end{tabular}
\label{tab:rates_bu}
\end{center}

\vspace*{.25cm} 

\begin{center}
\caption{Decay widths $\Lambda_c \to n \ell^+ \nu_\ell$ 
(in units of $|V_{cd}|^2$ ps$^{-1}$).} 

\vspace*{.15cm}

\def\arraystretch{1}
    \begin{tabular}{c|c|c}
      \hline
Mode & Our result  & Theoretical predictions \\ 
\hline
$\Lambda_c \to n e^+ \nu_e$ 
    & 0.20 
    & 0.26~\cite{Ivanov:1996fj}; 
      0.20~\cite{Pervin:2005ve};  
      0.27~\cite{Pervin:2005ve}\\
    & &
     $8.21 \pm 2.80$~\cite{Azizi:2009wn}; 
     $3.82 \pm 1.20$~\cite{Azizi:2009wn}; 
     $1.44 \pm 0.55$~\cite{Azizi:2009wn}; 
     $2.51 \pm 0.85$~\cite{Azizi:2009wn} \\
\hline
$\Lambda_c \to n \mu^+ \nu_\mu$ 
    & 0.19 
    & 0.26~\cite{Ivanov:1996fj}; 
      0.20~\cite{Pervin:2005ve};  
      0.27~\cite{Pervin:2005ve}\\ 
    & &
     $8.30 \pm 2.85$~\cite{Azizi:2009wn}; 
     $3.88 \pm 1.25$~\cite{Azizi:2009wn}; 
     $1.46 \pm 0.55$~\cite{Azizi:2009wn}; 
     $2.51 \pm 0.85$~\cite{Azizi:2009wn} \\
\hline
\end{tabular}
\label{tab:rates_cd}
\end{center}

\vspace*{.25cm}

\begin{center}
\caption{Asymmetry parameter $\alpha_{F\!B}^\ell$.} 

\vspace*{.15cm}

\def\arraystretch{1.25}
    \begin{tabular}{c|c}
      \hline
Mode & Our result  \\ 
\hline
$\Lambda_b \to p e^- \bar\nu_e$       & $-$ 0.388 \\
$\Lambda_b \to p \mu^- \bar\nu_\mu$   & $-$ 0.388 \\
$\Lambda_b \to p \tau^- \bar\nu_\tau$ & $-$ 0.434 \\
$\Lambda_c \to n e^+   \nu_e$         &     0.236 \\
$\Lambda_c \to n \mu^+ \nu_\mu$       &     0.209 \\
\hline
\end{tabular}
\label{tab:asym}
\end{center}
\end{table} 

\section{Summary} 

We have used the covariant constituent quark model previously developed
by us to calculate semileptonic heavy-to-light transitions of
$\Lambda_b$ and $\Lambda_c$ baryons. We have performed a detailed analysis
of the invariant and helicity amplitudes, form factors, angular decay
distributions, decay widths and asymmetry parameters. 
Following our 
previous papers~\cite{Faessler:2002ut,Kadeer:2005aq,Gutsche:2013pp,%
Gutsche:2013oea} we have used the helicity method in our analysis to 
provide complete 
information on the spin structure of the baryons and the off-shell $W$ 
boson. We have not provided an analysis of the polarization of the
charged lepton which, however, can be obtained in a straight-forward manner
using the helicity method as described in~\cite{Kadeer:2005aq}.
Our predictions will be useful for the ongoing experimental study 
of semileptonic heavy-to-light baryon decays. 

\begin{acknowledgments}

This work was supported by the DFG under Contract No. LY 114/2-1 
and by Tomsk State University Competitiveness Improvement Program.  
M.A.I.\ acknowledges the support from Mainz Institute for Theoretical 
Physics (MITP). M.A.I. and J.G.K. thank the Heisenberg-Landau Grant for
support.  

\end{acknowledgments}

\end{document}